\documentclass[aps,prl,twocolumn,superscriptaddress]{revtex4}

\newcommand{\ket}[1]{{|#1\rangle}} 

\begin{document}

\title{Comment on ``Probabilistic Quantum Memories''}

\author{T.Brun}
\affiliation{Electrical Engineering Dept., University of Southern California,
Los Angeles CA90089, USA}

\author{H.Klauck}
\affiliation{School of Mathematics, Institute for Advanced Study,
Princeton NJ08540, USA}

\author{A.Nayak} \author{M.R\"otteler} \author{Ch.Zalka}
\affiliation{Dept. of C\&O, University of Waterloo, Waterloo, Ontario,
Canada N2L 3G1}

\maketitle

In a series of similar articles \cite{t1,t2,t3} Trugenberger claims
that quantum states could be used as exponentially large memories for
classical information. This claim is wrong. Actually quantum mechanics
hardly offers any advantage for this task.

Trugenberger considers an ``associative memory'' in which
exponentially many binary strings are stored. For an additional binary
string, the goal is to find whether there are strings close to it (in
Hamming distance) in this memory. Also one would like to read out one
of these close strings. This might, e.g., be useful to find whether a
picture, given a noisy version of it, is in a large
database. Trugenberger proposes to use an $n$-qubit quantum register
as the memory. It is prepared in the uniform amplitude superposition
(eq. (3)) of exponentially many binary strings (here and in the
following we refer to \cite{t1}). Given the additional string, a
sequence of operations is performed which leads either to measurement
result $\ket{0}$ (= yes, similar patterns are in the memory) or
$\ket{1}$ (= no, there are no similar patterns). Also, if $\ket{0}$ is
measured, one of the similar strings in the memory is retrieved.

However it is easily seen that a simple classical scheme offers
exactly the same performance. Indeed we can replace the $n$-qubit
memory state with an $n$-bit classical memory which stores only a
single one of the binary strings. Consider the ``processed'' memory
state (eq. (16)) just before the measurement. In this state the
weights of all ``stored'' binary strings are still equal, thus no
amplification of states close to the additional string has taken
place. Because of this, we could simply classically store one of the
binary strings, chosen uniformly at random. Then we could compare this
random string with the additional one and, depending on how close they
are, decide to answer ``yes'' or ``no''. So the $n$ qubits can be
replaced with $n$ classical bits without changing the result. Of
course the performance of such a scheme is very poor, as it really
only stores a single bit string. Thus the repeated claims
\cite{t1,t2,t3} of exponential storage capacity, or actually of any
advantage over classical systems are wrong.

Furthermore the author wrongly assumes that this retrieval step could
relatively easily be repeated several times. He states that to this
end the ``memory state'' $\ket{M}$ could be cloned
probabilistically. In \cite{t3} he explains that this could be
achieved with a ``state dependent'' cloning machine. Note that a
retrieval mechanism which has to know about the data it is supposed to
retrieve, contradicts the idea of a memory, whether it is associative
or otherwise. In our case the cloning machine would have to know
virtually all the information about $\ket{M}$, namely all but possibly
$n$ bits, as a simple argument shows. Indeed to specify the state as
opposed to just the set of $2^n$ linearly independent states to which
it belongs (as proposed for the state dependent cloning scheme), one
needs at most an additional $n$ bits. Thus the advantage over a simple
re-preparation of $\ket{M}$ would be marginal (apart from the
disadvantage of it being probabilistic, which the author doesn't
discuss). In other words, we would really have to store the whole
database classically after all, contrary to the stated goal of the
scheme.

Actually it is known that for storing classical information, quantum
states in a certain sense cannot offer any advantage. In its simplest
version, for perfect channels, the Holevo bound \cite{holevo} states
that a quantum channel is no better than a classical one for
transmitting classical information (technically it is a bound on the
mutual information). This of course applies just as well to {\it
storing} classical information in quantum bits. A last possible
loophole might be ``quantum random access codes''. In these, we could
choose which one of a (possibly exponentially) large set of data sets
we want to retrieve from a quantum state (whereby the memory would
be destroyed). But at least in an asymptotic sense, i.e. large size
and high success probability, even this has been ruled out
\cite{nayak}. Thus we strongly doubt that quantum states could be
useful as memories for classical information.

\end{document}